\documentclass[11pt]{cernrep}
\usepackage{graphicx}
\def\ltap{\raisebox{-.4ex}{\rlap{$\sim$}} \raisebox{.4ex}{$<$}}
\def\gtap{\raisebox{-.4ex}{\rlap{$\sim$}} \raisebox{.4ex}{$>$}}
\newcommand {\pom} {I\!\!P}

\newcommand {\xpom} {x_{\pom}}
\begin{document}
\title{Diffraction for non-believers
\thanks{~~contributed to the
    Proceedings of the Workshop on HERA and the LHC, DESY and CERN,
    2004--2005.}
}
\author{Michele Arneodo$^{a}$ and Markus Diehl$^{b}$}
\institute{${}^{a}$Universit\`a del Piemonte Orientale, 28100 Novara,
  Italy \\ ${}^{b}$Deutsches Elektronen-Synchroton DESY, 22603
  Hamburg, Germany} 
\maketitle

\begin{abstract}
  Diffractive reactions involving a hard scale can be
  understood in terms of quarks and gluons.  These reactions have become a
  valuable tool for investigating the low-$x$ structure of the proton
  and the behavior of QCD in the high-density regime, and they may
  provide a clean environment to study or even discover the Higgs
  boson at the LHC.  In this paper we give a brief introduction to the
  description of diffraction in QCD.  We focus on key features studied
  in $ep$ collisions at HERA and outline challenges for understanding
  diffractive interactions at the LHC.
\end{abstract}

\pagenumbering{arabic}
\pagestyle{plain}

%%%%%%%%%%%%%%%%%%%%%%%%%%%%%%%%%%%%%%%%%%%%%%%%%%%%%%%%%%%%%

\section{Introduction}

\begin{figure}[hb]
\begin{center}
\includegraphics[width=3.4cm,angle=-90,clip=true]{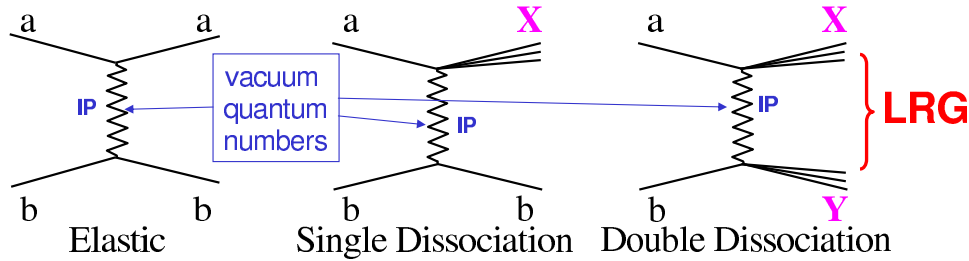}
\caption{Elastic scattering, single diffractive dissociation and
double diffractive dissociation in the collision of two hadrons 
$a$~and~$b$.  The two (groups of) final-state hadrons are separated by a 
large rapidity gap (LRG).  The zigzag lines denote the exchange of a 
Pomeron ($\pom$) in the $t$-channel. There are further graphs, not shown, 
with multiple Pomeron exchange.}
\label{diffraction_types} 
\end{center} 
\end{figure}

In hadron-hadron scattering a substantial fraction of the total cross
section is due to diffractive reactions.
Figure~\ref{diffraction_types} shows the different types of
diffractive processes in the collision of two hadrons: in elastic
scattering both projectiles emerge intact in the final state, whereas
single or double diffractive dissociation corresponds to one or both
of them being scattered into a low-mass state; the latter has the same
quantum numbers as the initial hadron and may be a resonance or
continuum state.  In all cases, the energy of the outgoing hadrons $a,
b$ or the states $X$, $Y$ is approximately equal to that of the
incoming beam particles, to within a few per cent.  The two (groups
of) final-state particles are well separated in phase space and in
particular have a large gap in rapidity between them.

Diffractive hadron-hadron scattering can be described within Regge
theory (see e.g.~\cite{Collins:1977jy}).  In this framework, the
exchange of particles in the $t$-channel is summed coherently to give
the exchange of so-called ``Regge trajectories''.  Diffraction is
characterized by the exchange of a specific trajectory, the
``Pomeron'', which has the quantum numbers of the vacuum.  Regge
theory has spawned a successful phenomenology of soft hadron-hadron
scattering at high energies.  Developed in the 1960s, it predates the
theory of the strong interactions, QCD, and is based on general
concepts such as dispersion relations.  Subsequently it was found that
QCD perturbation theory in the high-energy limit can be organized
following the general concepts of Regge theory; this framework is
often referred to as BFKL after the authors of the seminal papers
\cite{Kuraev:1976ge}.

It is clear that a $t$-channel exchange leading to a large rapidity gap 
in the final state must carry zero net color: if color were exchanged, 
the color field would lead to the production of further particles filling 
any would-be rapidity gap.  In QCD, Pomeron exchange is described by the 
exchange of two interacting gluons with the vacuum quantum numbers.

The effort to understand diffraction in QCD has received a great boost
from studies of diffractive events in $ep$ collisions at HERA (see 
e.g.~\cite{Forshaw:1997dc} for further reading and references).  The
essential results of these studies are discussed in the present paper
and can be summarized as follows:
\begin{itemize}
\item Many aspects of diffraction are well understood in QCD when a
  hard scale is present, which allows one to use perturbative
  techniques and thus to formulate the dynamics in terms of quarks and
  gluons.  By studying what happens when the hard
  scale is reduced towards the non-perturbative region, it may also be 
  possible to shed light on soft diffractive processes.
\item 
  Diffraction has become a tool to investigate low-momentum
  partons in the proton, notably through the study of diffractive
  parton densities in inclusive processes and of generalized parton
  distributions in exclusive ones.  
  Diffractive parton densities can be interpreted as conditional
  probabilities to find a parton in the proton when the final state of 
  the process contains a fast proton of given four-momentum.
  Generalized parton distributions, through their dependence on both 
  longitudinal and transverse variables, provide a three-dimensional 
  picture of the proton in high-energy reactions.
\item A fascinating link has emerged between diffraction and the
  physics of heavy-ion collisions through the concept of saturation,
  which offers a new window on QCD dynamics in the regime of high
  parton densities.
\end{itemize}
Perhaps unexpectedly, the production of the Higgs boson in diffractive 
$pp$ collisions is drawing more and more attention as a clean channel to 
study the properties of a light Higgs boson or even discover it. This is 
an example of a new theoretical challenge: to adapt and apply the 
techniques for the QCD description of diffraction in $ep$ collisions to 
the more complex case of $pp$ scattering at the LHC.  A first glimpse of 
phenomena to be expected there is provided by the studies of hard 
diffraction in $p\bar{p}$ collisions at the Tevatron.

%%%%%%%%%%%%%%%%%%%%%%%%%%%%%%%%%%%%%%%%%%%%%%%%%%%%%%%%%%%%%

\subsection{A digression on the nomenclature: why ``diffraction'' ?}
\label{optics}

\begin{figure}[tb]
\begin{center}
\includegraphics[width=7.2cm,angle=0]{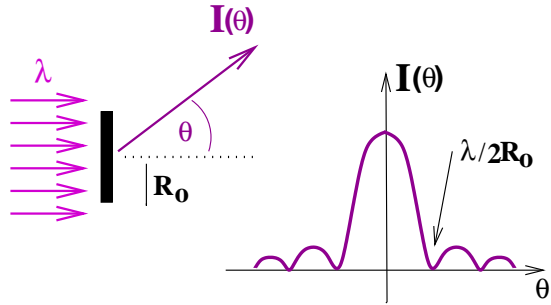}
\caption{Distribution of the intensity $I$ in the diffraction of light
of wavelength $\lambda$ from a circular target of size $R_0$.}
\label{diffraction-of-light} 
\end{center}
\end{figure}

Physics students first encounter the term ``diffraction'' in optics.
Light of wavelength $\lambda$ impinging on a black disk of radius
$R_0$ produces on a distant screen a diffraction pattern,
characterized by a large forward peak for scattering angle $\theta=0$
(the ``diffraction peak'') and a series of symmetric minima and
maxima, with the first minimum at $\theta_{\min}\simeq \pm
\lambda/(2R_0)$ (Fig.~\ref{diffraction-of-light}).  The intensity $I$
as a function of the scattering angle $\theta$ is given by
\begin{equation}
\frac{I(\theta)}{I(\theta=0)}=\frac{[2J_1(x)]^2}{x^2}
\simeq 1 - \frac{R_0^2}{4}(k\theta)^2,
\label{eq:light}
\end{equation}
where $J_1$ is the Bessel function of the first order and
$x=kR_0\sin{\theta} \simeq kR_0\, \theta$ with $k=2\pi/\lambda$.  The
diffraction pattern is thus related to the size of the target and to
the wavelength of the light beam.

\begin{figure}[tb]
\begin{center}
\includegraphics[width=11cm,angle=-90,clip=true]{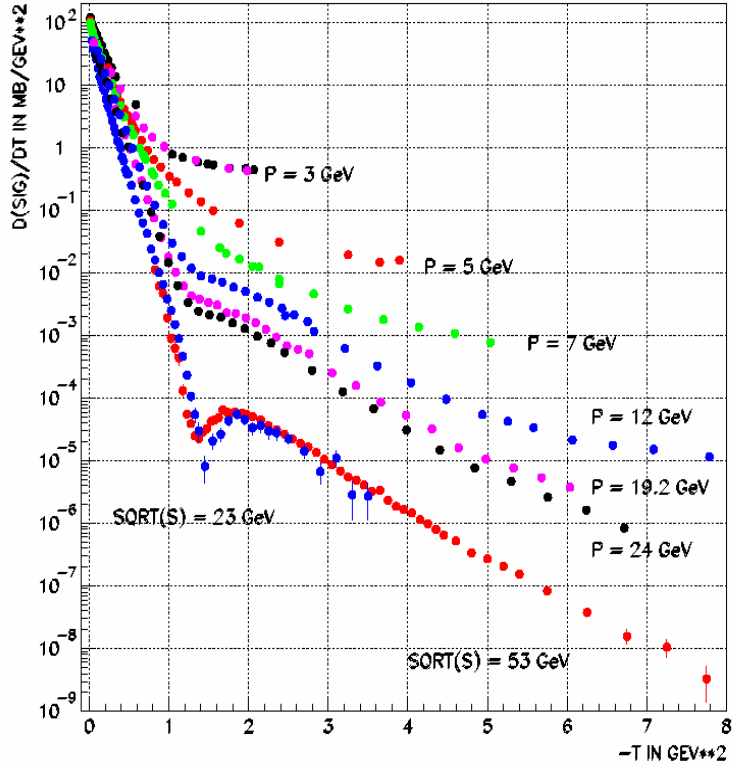}
\caption{Compilation of proton-proton elastic cross section data as a
  function of $t$.  The symbol $P$ indicates the momentum of the incoming  
  proton in a fixed target experiment and $\sqrt{s}$ the center-of-mass 
  energy in a $pp$ collider setup.}
\label{ppdsdt}
\end{center}
\end{figure}

As shown in Fig.~\ref{ppdsdt}, the differential cross section
$d\sigma/dt$ for elastic proton-proton scattering, $pp \rightarrow
pp$, bears a remarkable resemblance to the diffraction pattern just
described (see e.g. \cite{Goulianos:1982vk}).  At low values of $|t|$
one has
\begin{equation}
\frac{\frac{d\sigma}{dt}(t)}{\frac{d\sigma}{dt}(t=0)} 
\simeq e^{-b |t|}
\simeq 1 - b\, (P\theta)^2,
\label{eq:hadrons}
\end{equation}
where $|t| \simeq (P \theta)^2$ is the absolute value of the squared 
four-momentum transfer,
$P$ is the incident proton momentum and $\theta$ is the scattering
angle. The $t$-slope $b$ can be written as $b=R^2/4$, where once again
$R$ is related to the target size (or more precisely to the transverse
distance between projectile and target). A dip followed by a secondary
maximum has also been observed, with the value of $|t|$ at which the
dip appears decreasing with increasing proton momentum. It is hence 
not surprising that the term diffraction is used for elastic $pp$
scattering.  Similar $t$ distributions have been observed for the
other diffractive reactions mentioned above, leading to the use of the
term diffraction for all such processes.

%%%%%%%%%%%%%%%%%%%%%%%%%%%%%%%%%%%%%%%%%%%%%%%%%%%%%%%%%%%%%

\subsection{Diffraction at HERA ?!}
\label{diff-at-hera}

Significant progress in understanding diffraction has been made at the
$ep$ collider HERA, where 27.5~GeV electrons or positrons collide with
820 or 920~GeV protons. This may sound peculiar: diffraction is a
typical hadronic process while $ep$ scattering at HERA is an
electro-weak reaction, where the electron radiates a virtual photon
(or a $Z$ or $W$ boson), which then interacts with the
proton.\footnote{For simplicity we will speak of a virtual
photon in the following, keeping in mind that one can have a weak
gauge boson instead.}
To understand this, it is useful to look at $ep$ scattering in a frame
where the virtual photon moves very fast (for instance in the proton
rest frame, where the $\gamma^*$ has a momentum of up to about 50 TeV at
HERA).  The virtual photon can fluctuate into a quark-antiquark pair.
Because of its large Lorentz boost, this virtual pair has a lifetime
much longer than a typical strong interaction time.  In other words,
the photon fluctuates into a pair long before the collision, and it is
the pair that interacts with the proton. This pair is a small color
dipole. Since the interaction between the pair and the proton is
mediated by the strong interaction, diffractive events are possible.

An advantage of studying diffraction in $ep$ collisions is that, for
sufficiently large photon virtuality $Q^2$, the typical transverse
dimensions of the dipole are small compared to the size of a hadron.
Then the interaction between the quark and the antiquark, as well as the
interaction of the pair with the proton, can be treated perturbatively.
With decreasing $Q^2$ the color dipole becomes larger, and at very low
$Q^2$ these interactions become so strong that a description in terms
of quarks and gluons is no longer justified.  We may then regard the
photon as fluctuating into a vector meson -- this is the basis of the
well-known vector meson dominance model -- and can therefore expect to see
diffractive reactions very similar to those in hadron-hadron
scattering.

A different physical picture is obtained in a frame where the incident
proton is very fast.  Here, the diffractive reaction can be seen as the
deep inelastic scattering (DIS) of a virtual photon on the proton
target, with a very fast proton in the final state.  One can thus
expect to probe partons in the proton in a very specific way.  For
suitable diffractive processes there are in fact different types of
QCD factorization theorems, which bear out this expectation (see
Sects.~\ref{inclusive} and \ref{exclusive}).

%%%%%%%%%%%%%%%%%%%%%%%%%%%%%%%%%%%%%%%%%%%%%%%%%%%%%%%%%%%%%

\section{Inclusive diffractive scattering in \boldmath{$ep$}
  collisions} 
\label{inclusive}

\begin{figure}[tb]
\begin{center}
\includegraphics[width=0.45\textwidth]{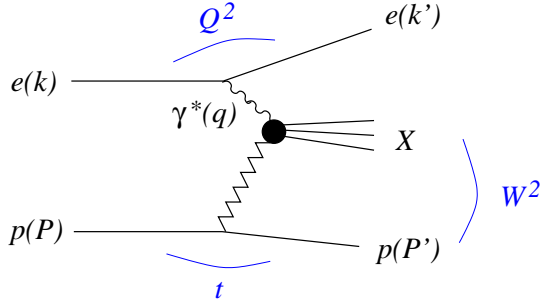}
\caption{Schematic diagram of inclusive diffractive DIS, $ep
  \rightarrow eXp$. Four-momenta are indicated in parentheses.}
\label{diffractive-dis-diagram}
\end{center}
\end{figure}

\begin{figure}[tb]
\begin{center}
\includegraphics[width=0.49\textwidth,bb=30 389 549 761]{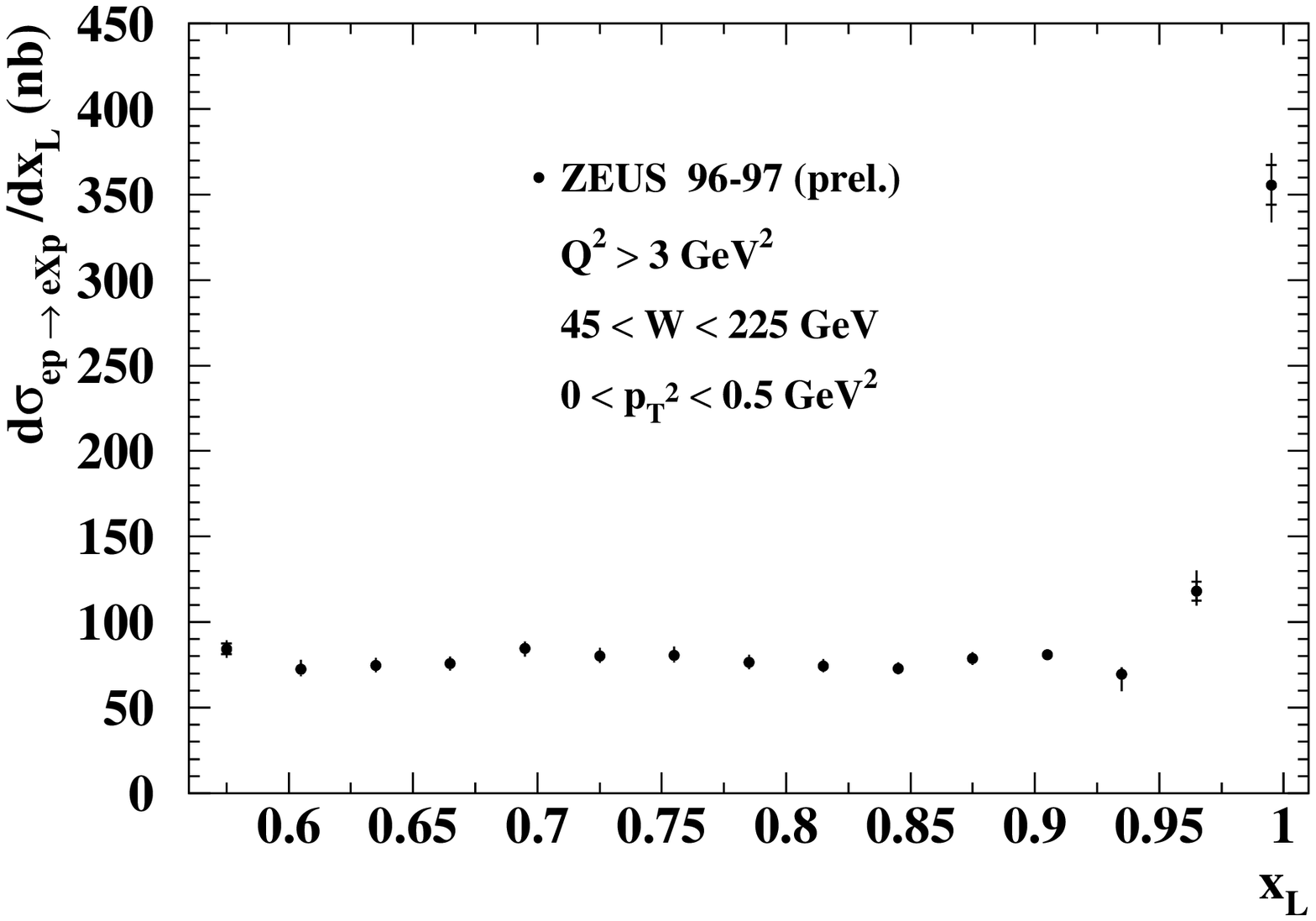}
\includegraphics[width=0.50\textwidth,bb=0 10 410 426]{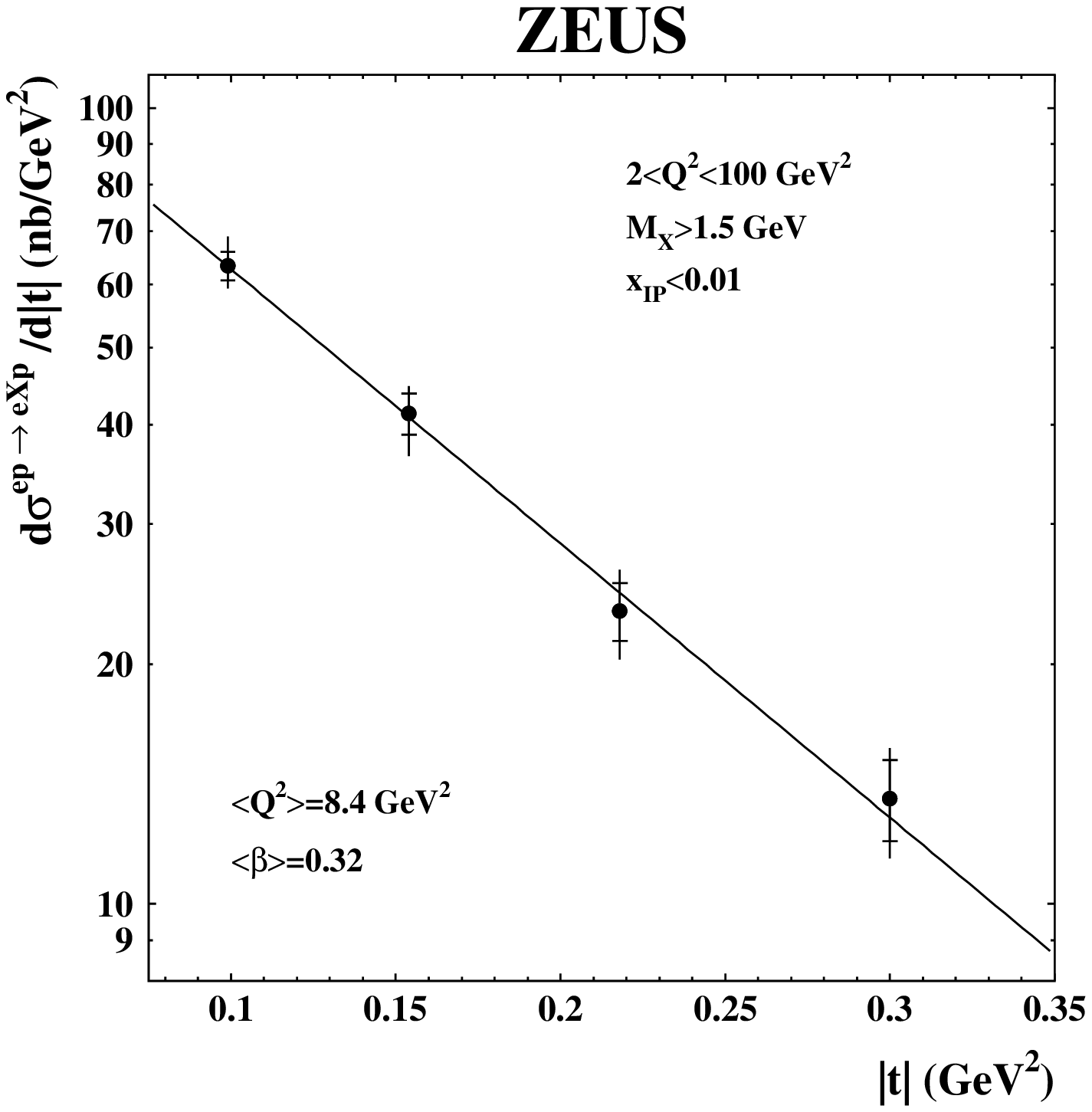}
\caption{Left: Differential cross section $d\sigma/dx_L$ for the
  process $ep \rightarrow eXp$ (from~\protect\cite{iacobucci}). The
  diffractive peak at $x_L \approx 1$ is clearly visible.  Right:
  Differential cross section $d\sigma/dt$ for the same process for
  $x_L> 0.99$ (from~\protect\cite{lps-f2d4}).  The average $|t|$ of
  this spectrum is $\langle |t| \rangle \approx 0.15$~GeV$^{2}$.}
\label{xlt_spectrum}
\end{center}
\end{figure}

\begin{figure}[tb]
\begin{center}
\includegraphics[width=6cm,angle=-90,clip=true]{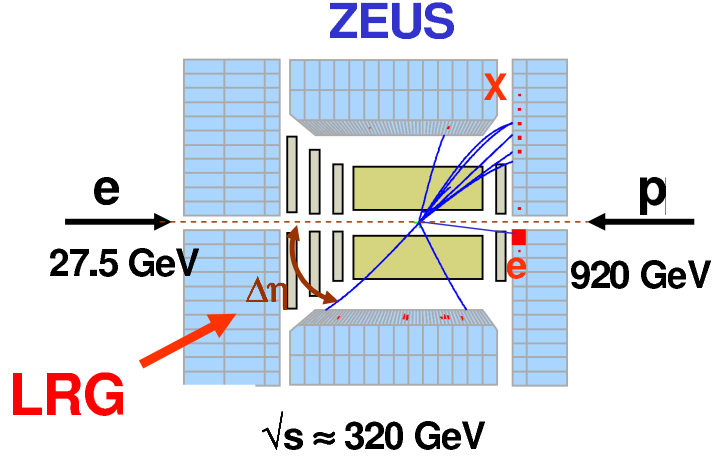}
\caption{A DIS event with a large rapidity gap (LRG) observed with the
  ZEUS detector at HERA.  The scattered proton escapes into the
  beam-pipe.  The symbol $\Delta \eta$ denotes the difference in
  pseudorapidity between the scattered proton and the most forward 
  particle of the observed hadronic system $X$. Pseudorapidity is defined 
  as  $\eta=-\ln{\tan({\theta/2})}$ in terms of the polar angle $\theta$
  measured with respect to the incoming proton direction, which is 
  defined as ``forward".}
\label{event_display}
\end{center}
\end{figure}

Figure~\ref{diffractive-dis-diagram} shows a schematic diagram of
inclusive diffractive DIS. The following features are important:
\begin{itemize}
\item The proton emerges from the interaction carrying a large
  fraction $x_L$ of the incoming proton momentum. Diffractive events
  thus appear as a peak at $x_L\approx 1$, the diffractive peak, which
  at HERA approximately covers the region $0.98<x_L<1$ (see the left
  panel of Fig.~\ref{xlt_spectrum}).  The right panel of 
  Fig.~\ref{xlt_spectrum} shows that large values of $|t|$ are 
  exponentially suppressed, similarly 
  to the case of elastic $pp$ scattering we discussed in Sect.~\ref{optics}.
  These protons remain in the beam-pipe and can only be measured with 
  detectors located inside the beam-pipe.
\item The collision of the virtual photon with the proton produces a
  hadronic final state $X$ with the photon quantum numbers and
  invariant mass $M_X$. A large gap in rapidity (or pseudorapidity) is
  present between $X$ and the final-state
  proton. Figure~\ref{event_display} shows a typical diffractive event
  at HERA.
\end{itemize}
Diffractive $ep$ scattering thus combines features of hard and soft
scattering.  The electron receives a large momentum transfer; in fact
$Q^2$ can be in the hundreds of GeV$^2$.  In contrast, the proton
emerges with its momentum barely changed.

%%%%%%%%%%%%%%%%%%%%%%%%%%%%%%%%%%%%%%%%%%%%%%%%%%%%%%%%%%%%%

\subsection{Diffractive structure functions}

The kinematics of $\gamma^* p \to Xp$ can be described by the
invariants $Q^2 =-q^2$ and $t =(P-P')^2$, and by the scaling variables
$x_{\pom}$ and $\beta$ given by
\begin{equation}
  \label{xpom-def}
x_{\pom} = \frac{(P-P')\cdot q}{P\cdot q} = 
\frac{Q^2+M_X^2-t}{W^2+Q^2-M_p^2} ,
\qquad
\beta = \frac{Q^2}{2(P-P')\cdot q} = \frac{Q^2}{Q^2+M_X^2-t} ,
\end{equation}
where $W^2 =(P+q)^2$ and the four-momenta are defined in
Fig.~\ref{diffractive-dis-diagram}. The variable $x_{\pom}$ is the
fractional momentum loss of the incident proton, related as $\xpom
\simeq 1-x_L$ to the variable $x_L$ introduced above.  The quantity
$\beta$ has the form of a Bjorken variable defined with respect to the
momentum $P-P'$ lost by the initial proton instead of the initial
proton momentum $P$. The usual Bjorken variable 
$x_B = Q^2 /(2 P\cdot q)$ is related to $\beta$ and $x_{\pom}$ as 
$\beta x_{\pom} = x_B$.

The cross section for $ep \to eXp$ in the one-photon exchange
approximation can be written in terms of diffractive structure
functions $F_2^{D(4)}$ and $F_L^{D(4)}$ as
\begin{equation}
\frac{d\sigma^{ep \rightarrow eXp}}{
d\beta\, dQ^2\,dx_{\pom}\,dt} = 
\frac{4\pi\alpha_{\mathrm{em}}^2}{\beta Q^4}
\biggl[ \Big(1-y+\frac{y^2}{2}\Big) F_2^{D(4)}(\beta,Q^2,x_{\pom},t) 
  - \frac{y^2}{2} F_L^{D(4)}(\beta,Q^2,x_{\pom},t) 
\biggr] ,
\label{sigma-2}
\end{equation}
in analogy with the way $d\sigma^{ep \rightarrow eX}/(dx_B\, dQ^2)$ 
is related to the structure functions $F_2$ and $F_L$ for inclusive DIS, 
$ep\to eX$. Here $y=(P\cdot q)/(P\cdot k)$ is the fraction of
energy lost by the incident lepton in the proton rest frame.  The
structure function $F_L^{D(4)}$ corresponds to longitudinal
polarization of the virtual photon; its contribution to the cross
section is small in a wide range of the experimentally accessible
kinematic region (in particular at low $y$).  The structure function
$F_2^{D(3)}$ is obtained from $F_2^{D(4)}$ by integrating over~$t$:
\begin{equation}
F_2^{D(3)}(\beta, Q^2,\xpom)=\int dt\,
F_2^{D(4)}(\beta,Q^2,x_{\pom},t)  .
\label{f2d3}
\end{equation}

\begin{figure}[t]
\begin{center}
\includegraphics[width=0.7\textwidth]{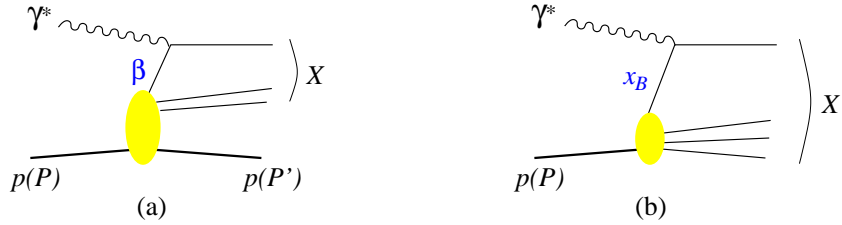}
\caption{Parton model diagrams for deep inelastic diffractive (a) and
  inclusive (b) scattering. The variable $\beta$ is the momentum
  fraction of the struck quark with respect to $P-P'$, and $x_B$ its
  momentum fraction with respect to $P$.}
\label{ddis-fact}
\end{center}
\end{figure}

In a parton model picture, inclusive diffraction $\gamma^* p \to Xp$
proceeds by the virtual photon scattering on a quark, in analogy to
inclusive scattering (see Fig.~\ref{ddis-fact}).  In this
picture, $\beta$ is the momentum fraction of the struck quark with
respect to the exchanged momentum $P-P'$ (indeed the allowed
kinematical range of $\beta$ is between 0 and 1).  The diffractive
structure function describes the proton structure in these specific
processes 
with a fast proton in the final state. $F_2^D$ may also be viewed as
describing the structure of whatever is exchanged in the $t$-channel
in diffraction, i.e. of the Pomeron (if multiple Pomeron exchange can
be neglected).  It is however important to bear in mind that the
Pomeron in QCD cannot be interpreted as a particle on which the
virtual photon scatters, as we will see in Sect.~\ref{notaparticle}.

\begin{figure}[p]
\begin{center}
\vspace{0.2cm}
\includegraphics[width=9.5cm,angle=-90,clip=true]{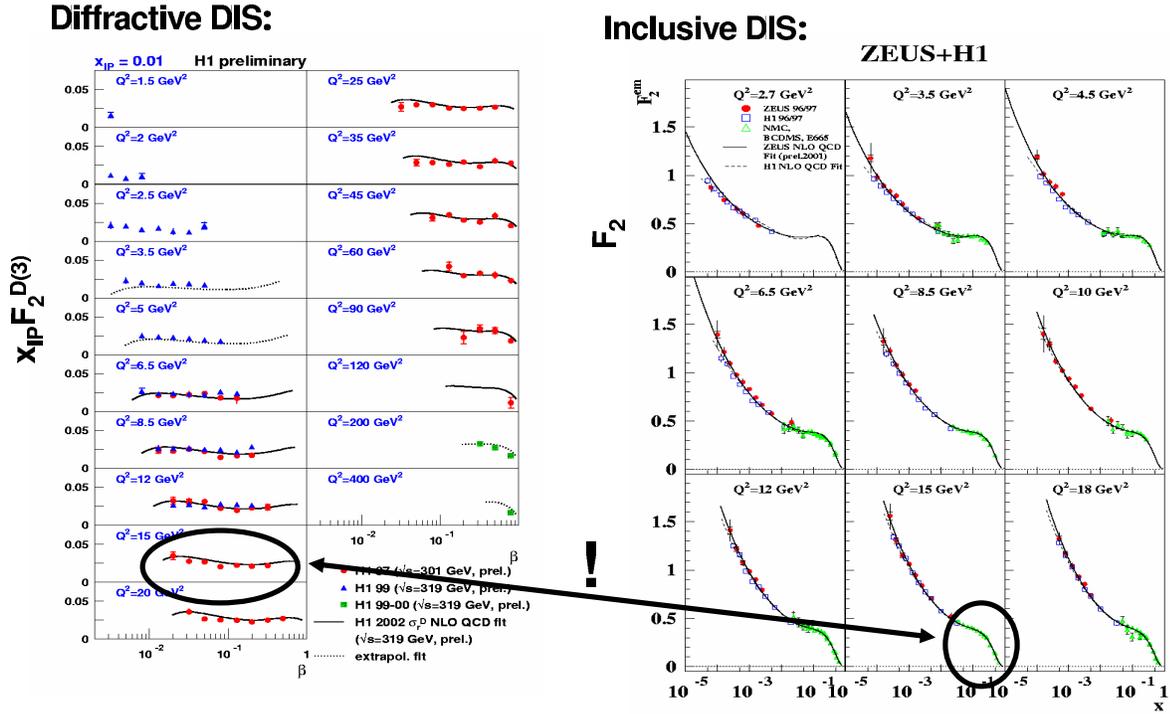}
\caption{Left: the diffractive structure function of the proton as a
  function of $\beta$ (from \protect\cite{H1prelim}). Right: the
  structure function of the proton as a function of $x_B$ (from
  \protect\cite{zeusf296-97prel}).  The two highlighted bins show 
  the different shapes of $F_2^D$ and $F_2$ in corresponding ranges of 
  $\beta$ and $x_B$ at equal $Q^2$.
}
\label{h1-fd2-vs-beta}
\end{center}
\end{figure}

\begin{figure}[p]
\begin{center}
\vspace{0.2cm}
\includegraphics[width=9.5cm,angle=-90,clip=true]{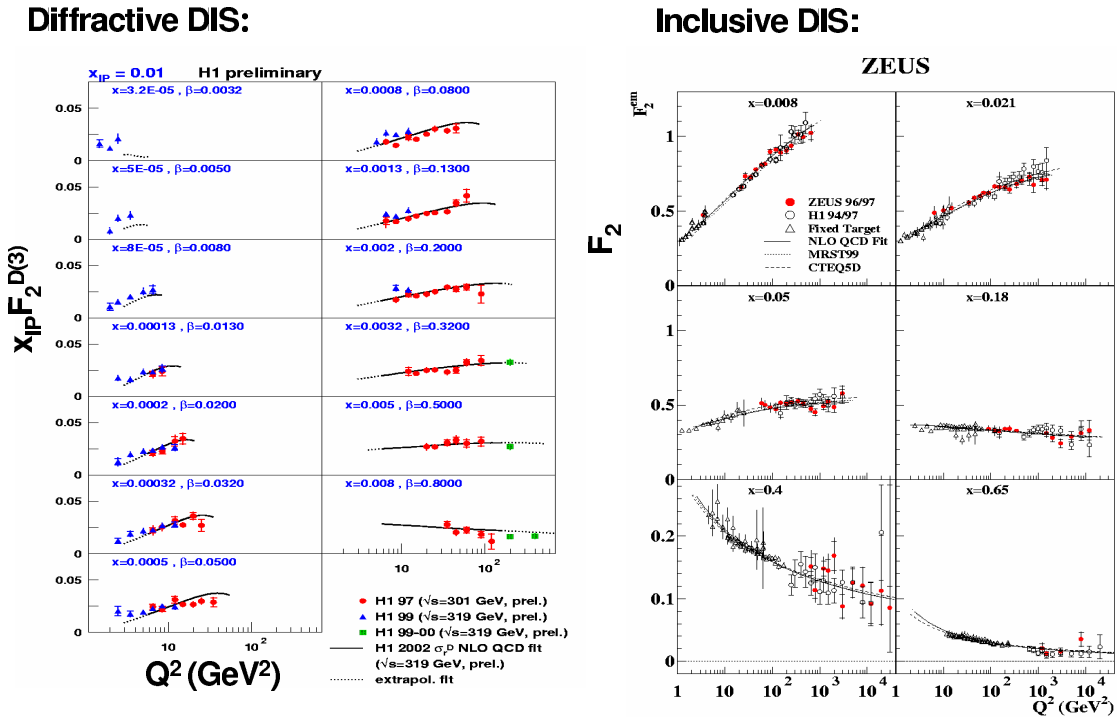}
\caption{Left: the diffractive structure function of the proton as a
  function of $Q^2$ (from \protect\cite{H1prelim}). Right: the
  structure function of the proton as a function of $Q^2$ (from
  \protect\cite{zeusf296-97}).}
\label{h1-fd2-vs-q2}
\end{center}
\end{figure}

Figures~\ref{h1-fd2-vs-beta} and \ref{h1-fd2-vs-q2} show recent H1
data~\cite{H1prelim} on $F_2^{D(3)}$ at fixed $\xpom$ as a function of
$\beta$ for different $Q^2$ bins, and as a function of $Q^2$ for
different bins of $\beta$.\footnote{To be precise, the H1 data are for 
the so-called reduced diffractive cross section, which equals $F_2^{D(3)}$ 
if $F_L^D$ can be neglected.}  The data have two remarkable 
features:
\begin{itemize}
\item 
  $F_2^D$ is largely flat in the measured $\beta$ range. Keeping in
  mind the analogy between $\beta$ in diffractive DIS and $x_B$ in
  inclusive DIS, this is very different from the behavior of the
  ``usual'' structure function $F_2$, which strongly decreases for
  $x_B \;\gtap\; 0.2$ (see Fig.~\ref{h1-fd2-vs-beta}).
\item The dependence on $Q^2$ is logarithmic, i.e. one observes
  approximate Bjorken scaling.  This indicates the applicability of
  the parton model picture to inclusive $\gamma^* p$ diffraction.  The
  structure function $F_2^D$ increases with $Q^2$ for all $\beta$
  values except the highest. This is reminiscent of the scaling
  violations of $F_2$, except that $F_2$ rises with $Q^2$ only for
  $x_B \;\ltap\; 0.2$ and that the scaling violations become negative
  at higher $x_B$ (see Fig.~\ref{h1-fd2-vs-q2}).  In the proton,
  negative scaling violations reflect the presence of the valence
  quarks radiating gluons, while positive scaling violations are due
  to the increase of the sea quark and gluon densities as the proton
  is probed with higher resolution. The $F_2^D$ data thus suggest that
  the partons resolved in diffractive events are predominantly gluons.
  This is not too surprising if one bears in mind that these partons
  carry only a small part of the proton momentum: the struck quark in
  the diagram of Fig.~\ref{ddis-fact}a has a momentum fraction
  $\beta x_{\pom} = x_B$ with respect to the incident proton, and
  $\xpom \;\ltap\; 0.02$~--~$0.03$ in diffractive events.
\end{itemize}

%%%%%%%%%%%%%%%%%%%%%%%%%%%%%%%%%%%%%%%%%%%%%%%%%%%%%%%%%%%%%

\subsection{Diffractive parton distributions}
\label{dpdfs}

The conclusion just reached can be made quantitative by using the QCD
factorization theorem for inclusive diffraction, $\gamma^* p\to Xp$,
which formalizes the parton model picture we have already invoked 
in our discussion.  According to this theorem, the diffractive
structure function, in the limit of large $Q^2$ at fixed $\beta$,
$\xpom$ and $t$, can be written as
\cite{Trentadue:1993ka,Berera:1995fj,Collins:1997sr}
\begin{equation}
  \label{fact-theo}
F_{2}^{D(4)}(\beta,Q^2,x_{\pom},t) =
 \sum_i \int_{\beta}^1 \frac{dz}{z}\,
     C_i\Big(\frac{\beta}{z}\Big)\, f_i^D(z,x_{\pom},t;Q^2) ,
\end{equation}
where the sum is over partons of type $i$.  The coefficient functions
$C_i$ describe the scattering of the virtual photon on the parton and
are exactly the same as in inclusive DIS.  
In analogy to the usual parton distribution functions (PDFs), the
diffractive PDFs $f_i^D(z,\xpom,t;Q^2)$ can be defined as operator
matrix elements in a proton state, and their dependence on the
scale $Q^2$ is given by the DGLAP evolution equations.
In parton model language, they can be interpreted as conditional
probabilities  to find a parton $i$ with fractional momentum $z \xpom$ in 
a proton, probed with resolution $Q^2$ in a process with a fast proton in 
the final state (whose momentum is specified by $\xpom$ and $t$).

\begin{figure}[p]
\begin{center}
\includegraphics[width=11cm,angle=0]{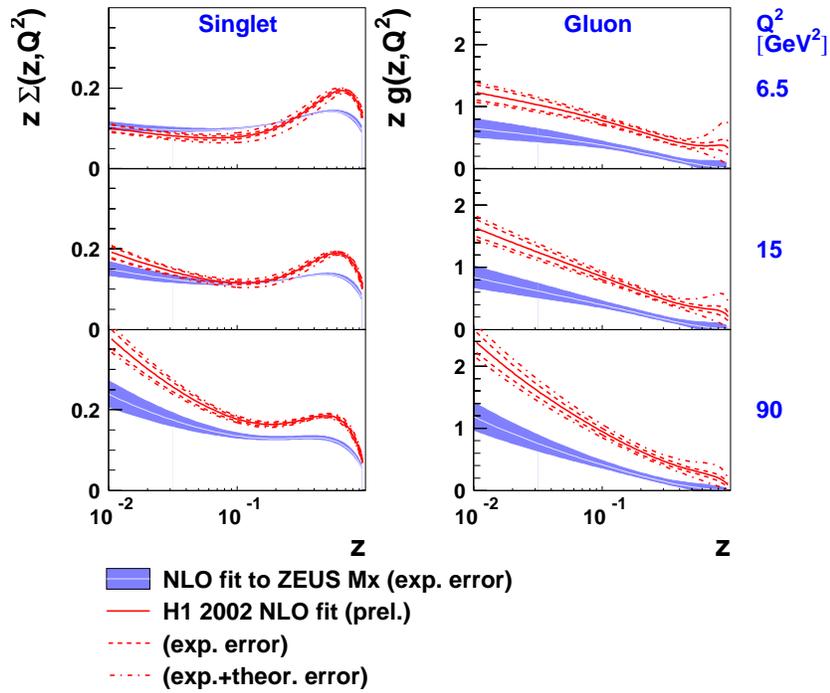}
\caption{Diffractive quark singlet and gluon distributions obtained
  from fits to H1~\protect\cite{H1prelim} and
  ZEUS~\protect\cite{Chekanov:2005vv} data
  (from~\protect\cite{schilling}).}
\label{fig-dpdfs}
\end{center}
\end{figure}

During the workshop, several fits of the available $F_2^D$ data 
were discussed which are based on the factorization formula
(\ref{fact-theo}) at next-to-leading order (NLO) in $\alpha_s$
\cite{schilling,glp}.  Figure~\ref{fig-dpdfs} compares the diffractive
PDFs from an earlier H1 fit~\cite{H1prelim} to those from the fit of
the ZEUS data~\cite{Chekanov:2005vv} by Schilling and
Newman~\cite{schilling}.  As expected the density of gluons is larger
than that of quarks, by about a factor 5--10.  Discrepancies between the
two sets are evident, and it remains to be clarified to which extent
they reflect differences in the fitted data.  Martin, Ryskin and
Watt~\cite{watt} have argued that the leading-twist formula
(\ref{fact-theo}) is inadequate in large parts of the measured
kinematics, and performed a fit to a modified expression which
includes an estimate of power-suppressed effects.  The discrepancies
between the various diffractive PDFs, while not fully understood, may
be taken as an estimate of the uncertainties on these functions at
this point in time.  A precise and consistent determination of the
diffractive PDFs and their uncertainties is one of the main tasks the HERA 
community has to face in the near future.  They are a crucial input for 
predicting cross sections of inclusive diffractive processes at the LHC.

%%%%%%%%%%%%%%%%%%%%%%%%%%%%%%%%%%%%%%%%%%%%%%%%%%%%%%%%%%%%%

\subsection{Diffractive hard-scattering factorization}
\label{diff-fact}

\begin{figure}[p]
\begin{center}
\includegraphics[width=12cm,angle=0]{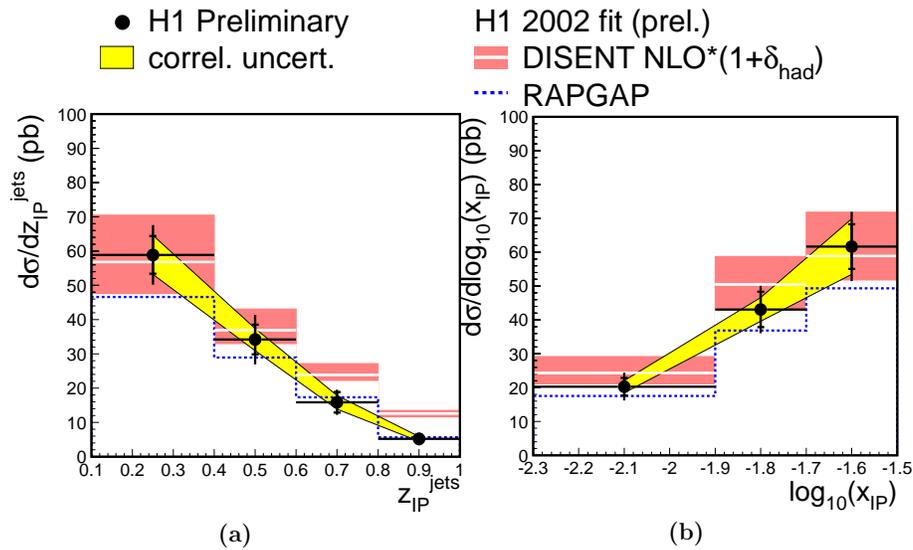}
\caption{Cross section for dijet production in diffractive DIS,
  compared with the expectations based on the diffractive
  PDFs~\protect\cite{H1prelim} (from~\protect\cite{H1-prelim-dijets}).
  The variable $z_{\pom}^{\rm jets}$ estimates the fractional momentum
  of the parton entering the hard subprocess.  }
\label{h1-dijets}
\end{center}
\end{figure}

Like usual parton densities, diffractive PDFs are process-independent
functions.  They appear not only in inclusive diffraction but also in
other processes where diffractive hard-scattering factorization holds.  
 In analogy with
  Eq.~(\ref{fact-theo}), the cross section of such a process can be
  evaluated as the convolution of the relevant parton-level
  cross section with the diffractive PDFs.
For instance, the cross section
for charm production in diffractive DIS can be calculated at leading order
in $\alpha_s$ from the $\gamma^* g \rightarrow c \bar c$ cross section and
the diffractive gluon distribution.  An analogous statement holds for jet
production in diffractive DIS. Both processes have been analyzed at
next-to-leading order in $\alpha_s$.

As an example, Fig.~\ref{h1-dijets} shows a comparison between the
measured cross sections for diffractive dijet production and the
expectations based on diffractive PDFs extracted from a fit to
$F_2^D$. These data lend support to the validity of hard-scattering
factorization in diffractive $\gamma^* p$ interactions.  For further
discussion we refer the reader to~\cite{alessia}.

%%%%%%%%%%%%%%%%%%%%%%%%%%%%%%%%%%%%%%%%%%%%%%%%%%%%%%%%%%%%%

\subsection{Limits of diffractive hard-scattering factorization:
hadron-hadron collisions}
\label{survival}

\begin{figure}[h]
\begin{center}
\includegraphics[width=9cm,bb=0 15 465 465]{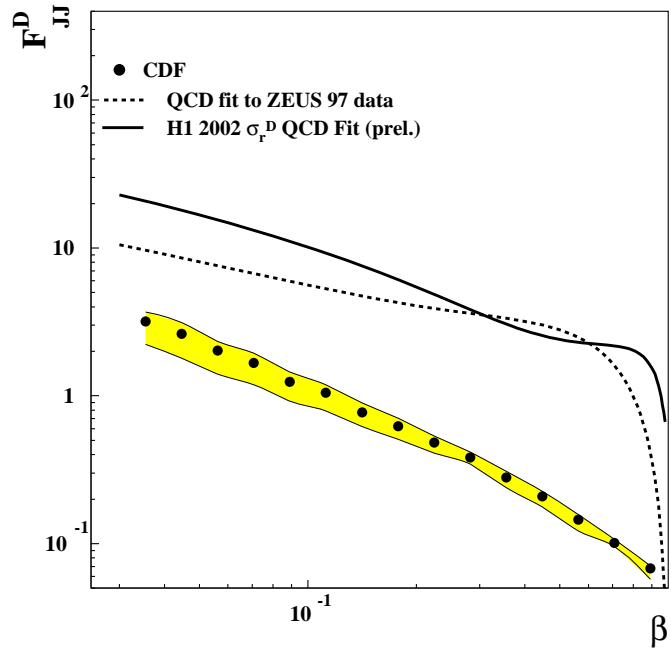}
\caption{CDF results for the cross section of diffractive dijet
  production with a leading antiproton in $p\bar{p}$ collisions
  (expressed in terms of a structure function $F^D_{JJ}$), compared with
  the predictions obtained from the diffractive
  PDFs~\protect\cite{lps-f2d4} and~\protect\cite{H1prelim} extracted
  at HERA (from~\protect\cite{arneodo}). See also the analogous plot in 
the original CDF publication~\protect\cite{cdf}.}
\label{fig-cdf}
\end{center}
\end{figure}

A natural question to ask is whether one can use the diffractive PDFs
extracted at HERA to describe hard diffractive processes such as the
production of jets, heavy quarks or weak gauge bosons in $p\bar{p}$
collisions at the Tevatron.  Figure~\ref{fig-cdf} shows results on
diffractive dijet production from the CDF collaboration~\cite{cdf}
compared to the expectations based on the diffractive
PDFs~\cite{lps-f2d4,H1prelim} from HERA.  The discrepancy is spectacular:
the fraction of diffractive dijet events at CDF is a factor 3 to 10
smaller than would be expected on the basis of the HERA data. The same
type of discrepancy is consistently observed in all hard diffractive
processes in $p\bar{p}$ events, see e.g.~\cite{Alvero:1998ta}.  In
general, while at HERA hard diffraction contributes a fraction of order
10\% to the total cross section, it contributes only about 1\% at the
Tevatron.

\begin{figure}[t]
\begin{center}
\includegraphics[width=0.4\textwidth]{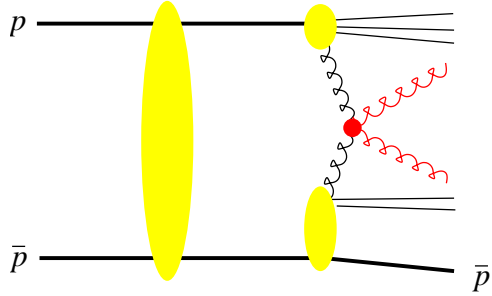}
\caption{Example graph for diffractive dijet production with a leading
  antiproton in a $p\bar{p}$ collision.  The interaction indicated by
  the large vertical blob breaks hard diffractive factorization. It 
  reduces the diffractive cross section, as explained in the text.}
\label{fact-break}
\end{center}
\end{figure}

In fact, diffractive hard-scattering factorization does not apply to
hadron-hadron collisions~\cite{Berera:1995fj,Collins:1997sr}.
Attempts to establish corresponding factorization theorems fail
 because of interactions between spectator partons of the colliding
   hadrons.  The contribution of these interactions to the cross section
   does not decrease with the hard scale.  Since they are not associated
   with the hard-scattering subprocess (see Fig.~\ref{fact-break}), we no
   longer have factorization into a parton-level cross section and the
   parton densities of one of the colliding hadrons. These
interactions are generally soft, and we have at present to rely on
phenomenological models to quantify their effects \cite{maor}.

The yield of diffractive events in hadron-hadron collisions is lowered
precisely because of these soft interactions between spectator partons
(often referred to as ``reinteractions'' or ``multiple scatterings'').  
They can produce additional final-state particles which fill the would-be
rapidity gap (hence the often-used term ``rapidity gap survival'').  When
such additional particles are produced, a very fast proton can no longer
appear in the final state because of energy conservation.  Diffractive
factorization breaking is thus intimately related to multiple scattering
in hadron-hadron collisions; understanding and describing this
phenomenon is a challenge in the high-energy regime that will be reached
at the LHC~\cite{multiple}.

In $pp$ or $p\bar{p}$ reactions, the collision partners are both
composite systems of large transverse size, and it is not too
surprising that multiple interactions between their constituents can
be substantial.  In contrast, the virtual photon in $\gamma^* p$
collisions has small transverse size, which disfavors multiple
interactions and enables diffractive factorization to hold.  According
to our discussion in Sect.~\ref{diff-at-hera}, we may expect that for
decreasing virtuality $Q^2$ the photon behaves more and more like a
hadron, and diffractive factorization may again be broken.
This aspect of diffractive processes in photoproduction at HERA was 
intensively discussed during the workshop, and findings are
reported in~\cite{alessia}.

%%%%%%%%%%%%%%%%%%%%%%%%%%%%%%%%%%%%%%%%%%%%%%%%%%%%%%%%%%%%%

\subsection{Space-time structure: the Pomeron is not a particle}
\label{notaparticle}

It is tempting to interpret diffractive $\gamma^* p$ processes as the
scattering of a virtual photon on a Pomeron which has been radiated
off the initial proton.  Diffractive DIS would then probe the
distribution of partons in a ``Pomeron target''.  This is indeed the
picture proposed by Ingelman and Schlein long
ago~\cite{Ingelman:1984ns}.

\begin{figure}[b]
\begin{center}
\includegraphics[width=1\textwidth]{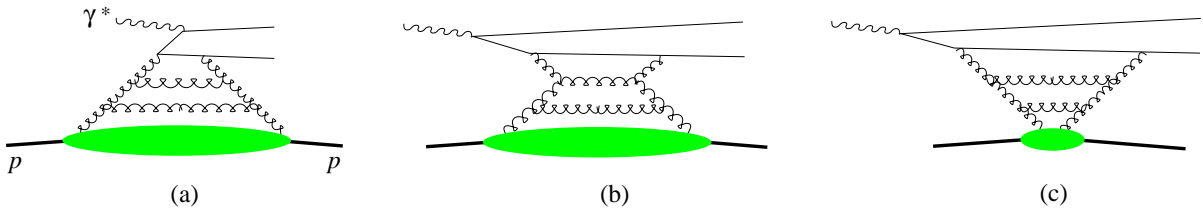}
\caption{Dominant time ordering for diffractive dissociation of a
  virtual photon in (a) the Breit frame, (b) the photon-proton
  center-of-mass, (c) the proton rest frame.  The physical picture in
  (a) corresponds to the parton-model description of diffraction, and
  the one in (b) and (c) to the picture of the photon splitting into a
  quark-antiquark dipole which subsequently interacts with the  
  proton.}
\label{time-ordering}
\end{center}
\end{figure}

This picture is however not supported by an analysis in QCD (see
e.g.~\cite{Bartels:1996hw}). There, high-energy scattering is dominated by
the exchange of two gluons, whose interaction is (in an appropriate gauge)
described by ladder diagrams, as shown in Fig.~\ref{time-ordering}. By
analyzing these diagrams in time-ordered perturbation theory, one can
obtain the dominant space-time ordering in the high-energy limit.  The
result depends on the reference frame, as illustrated in the figure.  In
the Breit frame, which is natural for a parton-model interpretation, the
photon does \emph{not} scatter off a parton in a pre-existing two-gluon
system; in fact some of the interactions in the gluon ladder building up
the Pomeron exchange take place long after the virtual photon has been
absorbed.  The picture in the Breit frame is however compatible with the
interpretation of diffractive parton densities given in Sect.~\ref{dpdfs},
namely the probability to find a parton under the condition that
subsequent interactions will produce a fast proton in the final state.

We note that the Ingelman-Schlein picture suggests that the
diffractive structure function takes a factorized form $F_2^{D(4)} =
f_{\pom}(\xpom,t)\, F_2^{\pom}(\beta,Q^2)$, where $f_{\pom}$ is the   
``Pomeron flux'' describing
  the emission of the Pomeron from the proton and its subsequent
propagation, and where $F_2^{\pom}$ is the ``structure function of the
Pomeron''.  Phenomenologically, such a factorizing ansatz works not too
badly and is often used, but recent high-precision data have shown its
breakdown at small $\xpom$~\cite{Chekanov:2005vv}.

%%%%%%%%%%%%%%%%%%%%%%%%%%%%%%%%%%%%%%%%%%%%%%%%%%%%%%%%%%%%%

\section{Exclusive diffractive processes}
\label{exclusive}

\begin{figure}[htb]
\begin{center}
\includegraphics[width=0.65\textwidth]{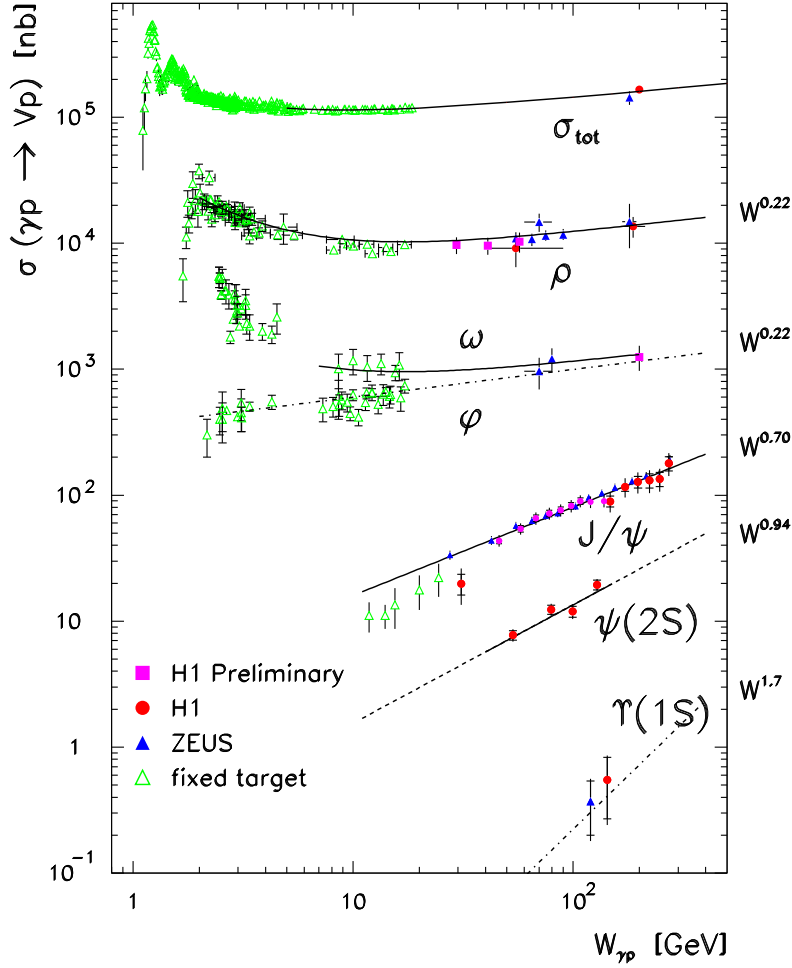}
\caption{\label{vmXsec} Compilation of results on the cross section
  for vector meson photoproduction, $\gamma p \rightarrow Vp$, with
  $V=\rho$, $\omega$, $\phi$, $J/\Psi$, $\psi'$, $\Upsilon$, as a
  function $W$.  The total $\gamma p$ cross section
  $\sigma_{\mathrm{tot}}$ is also shown.}
\end{center}
\end{figure}

\begin{figure}[htb]
\begin{center}
\includegraphics[width=0.6\textwidth,bb=20 250 545 760,%
  clip=true]{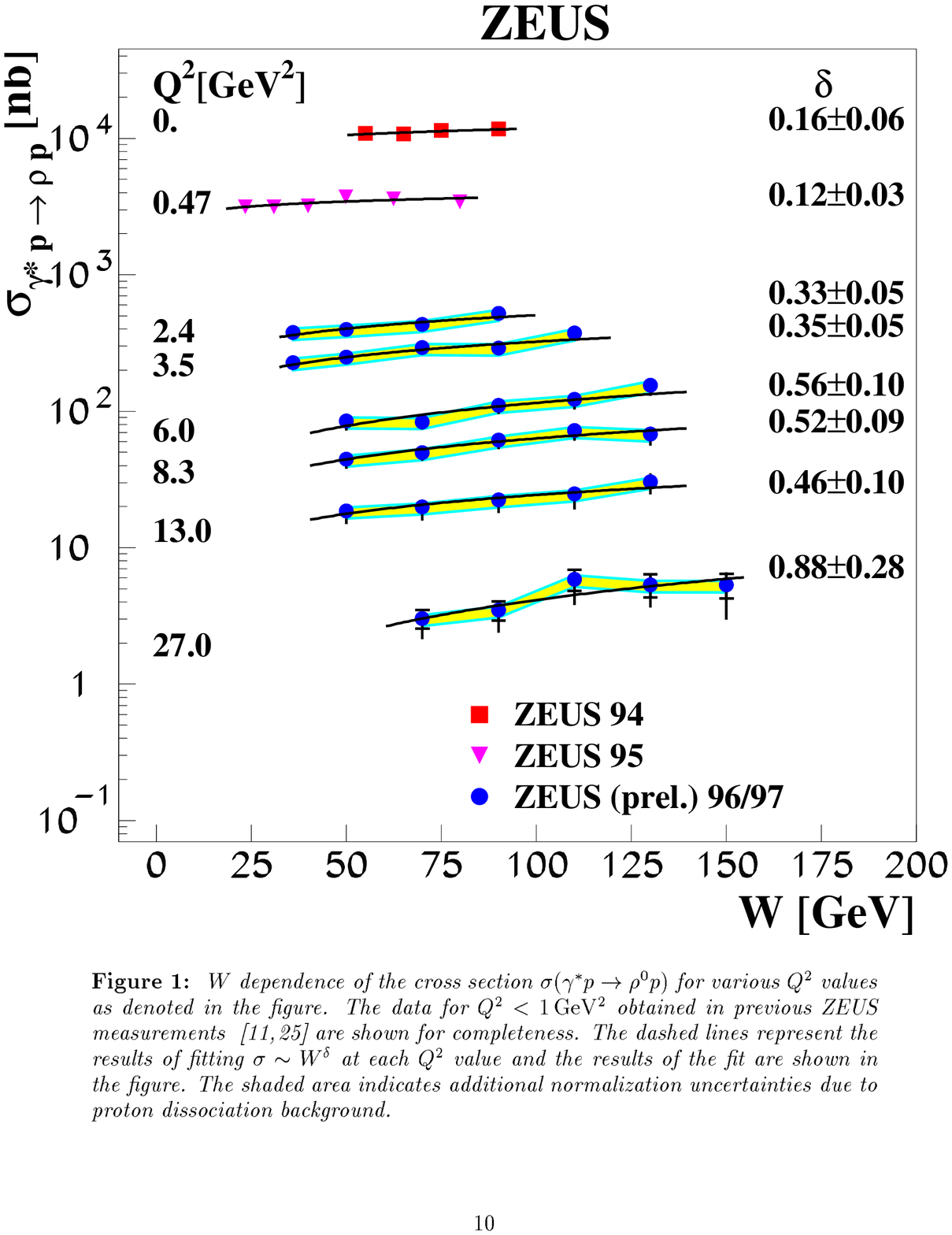}
\caption{\label{rhoXsec} Cross section for exclusive $\rho$ production
  as a function of $W$ (from \protect\cite{zeus-rho}). The lines
  represent the result of fits to the data with the form
  $\sigma(\gamma^*p\to \rho p)\propto W^\delta$, yielding the
  exponents given in the figure.}
\end{center}
\end{figure}

Let us now discuss diffractive processes where a real or virtual
photon dissociates into a single particle.  Since diffraction involves
the exchange of vacuum quantum numbers, this particle can in
particular be a vector meson (which has the same $J^{PC}$ quantum
numbers as the photon) -- in this case the process is sometimes referred 
to as ``elastic'' vector meson production.  Another important case is 
deeply virtual Compton scattering (DVCS), $\gamma^*p \to \gamma
p$.\footnote{We do not discuss processes with diffractive dissociation
of the proton in this paper, but wish to mention interesting studies
of vector meson or real photon production at large $|t|$, where the
proton predominantly dissociates, see e.g.~\protect\cite{Enberg:2004jv}.}
A striking feature of the data taken at HERA (Figs.~\ref{vmXsec} and
\ref{rhoXsec}) is that the energy dependence of these processes
becomes steep in the presence of a hard scale, which can be either the
photon virtuality $Q^2$ or the mass of the meson in the case of
$J/\Psi$ or $\Upsilon$ production.  This is similar to the energy
dependence of the $\gamma^* p$ total cross section (related by the
optical theorem to forward Compton scattering, $\gamma^* p\to \gamma^*
p$), which changes from flat to steep when going from real photons to
$Q^2$ of a few GeV$^2$.

To understand this similarity, let us recall that in perturbative QCD
diffraction proceeds by two-gluon exchange.  The transition from a
virtual photon to a real photon or to a quark-antiquark pair
subsequently hadronizing into a meson is a short-distance process 
involving these gluons,
provided that either $Q^2$ or the quark mass is large.  In fact, in an 
approximation discussed below, the cross sections for DVCS and vector 
meson
production are proportional to the square of the gluon distribution in
the proton, evaluated at a scale of order $Q^2+M_V^2$ and at a momentum
fraction $\xpom= (Q^2+M_V^2) /(W^2+Q^2)$, where the vector meson mass
$M_V$ now takes the role of $M_X$ in inclusive diffraction
\cite{Ryskin:1992ui}. In analogy to the case of the total $\gamma^* p$ 
cross section, 
the energy dependence of the cross sections shown in Figs.~\ref{vmXsec} 
and \ref{rhoXsec} thus reflects the $x$
and scale dependence of the gluon density in the proton, which grows
with decreasing $x$ with a slope becoming steeper as the scale
increases.

\begin{figure}[p]
\begin{center}
\includegraphics[width=12cm]{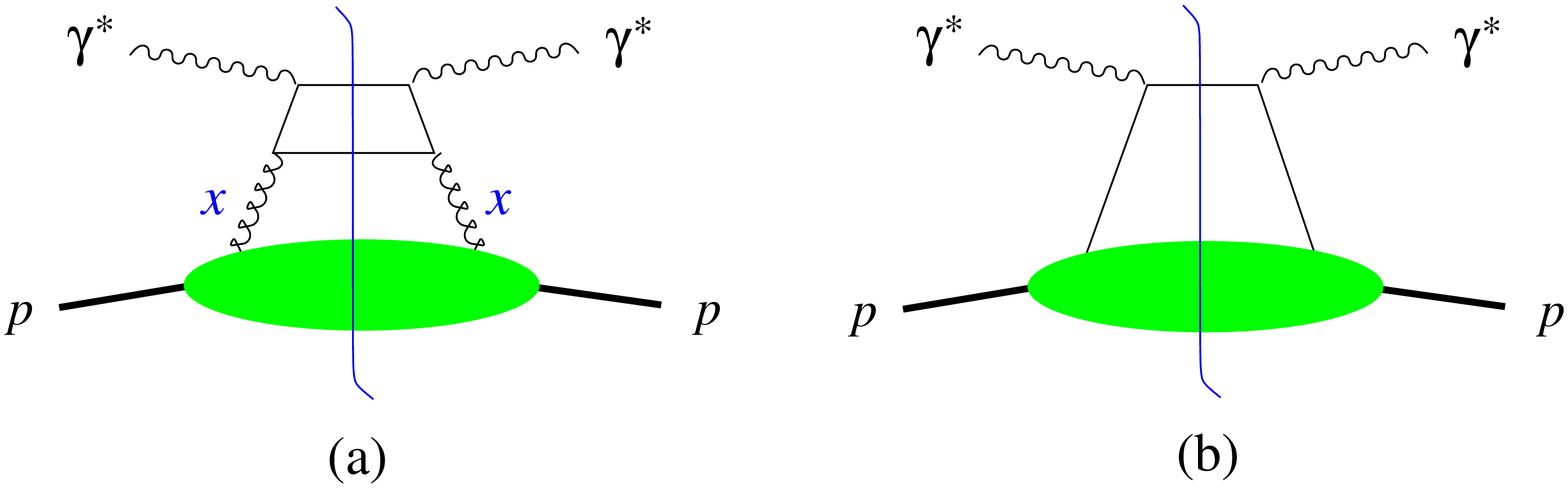}
\caption{\label{compton-fact} Factorization of forward Compton
    scattering, which is related to the total inclusive structure
    function via the optical theorem, $\mathrm{Im}\,
    \mathcal{A}(\gamma^* p\to\gamma^* p) = \frac{1}{2} \sum_X |
    \mathcal{A}(\gamma^* p\to X) |^2 \propto \sigma(\gamma^*p\to X)$.
    The final state of the inclusive process is obtained by cutting the
    diagrams along the vertical line. The blobs represent the gluon or 
    quark distribution in the proton. 
    Graph (b) is absent in the $k_t$ factorization formalism (see 
    Sect.~\protect\ref{sec:dipole-pic}): its role is taken by graph (a) in the
    ``aligned jet configuration'', where the quark line joining the
    two photons carries almost the entire photon momentum.}
\end{center}
\end{figure}

\begin{figure}[p]
\begin{center}
\includegraphics[width=12cm]{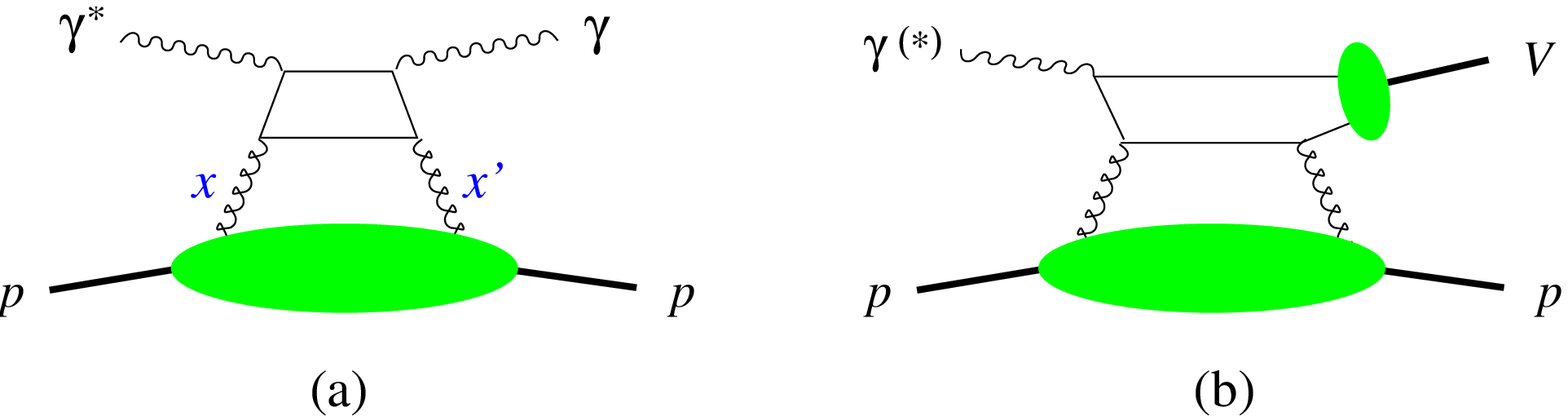}
\caption{\label{meson-fact} (a) Factorization of deeply virtual
  Compton scattering, $\gamma^* p\to \gamma p$, which can be measured
  in the exclusive process $ep\to ep \gamma$.  The blob represents 
  the generalized gluon distribution, with $x$ and
  $x'$ denoting the momentum fractions of the gluons. 
  (b) Factorization of exclusive meson production.  The small blob represents 
  the vector meson wave function. In the
  collinear factorization formalism, there are further graphs (not
  shown) involving quark instead of gluon exchange. 
}
\end{center}
\end{figure}

There is however an important difference in how the gluon distribution 
enters
the descriptions of inclusive DIS and of exclusive diffractive
processes.  The inclusive DIS cross section is related via the optical
theorem to the imaginary part of the forward virtual Compton
amplitude, so that the graphs in Fig.~\ref{compton-fact} represent
the \emph{cross section} of the inclusive process. Hence, the 
gluon distribution in Fig.~\ref{compton-fact}a gives the 
\emph{probability} to find \emph{one} gluon 
in the proton (with any number of unobserved spectator partons going into
the final state).  
 In contrast, the corresponding graphs for DVCS and exclusive meson
 production in Fig.~\ref{meson-fact} represent the \emph{amplitudes} of
 exclusive processes, which are proportional to the \emph{probability
 amplitude} for first extracting a gluon from the initial proton and then
 returning it to form the proton in the final state.  In the approximation 
 discussed below, this probability amplitude is given by the gluon
 distribution.  The cross sections of DVCS and exclusive meson production
 are then proportional to the \emph{square} of the gluon distribution.

A detailed theoretical analysis  of DVCS and exclusive meson
production at large $Q^2$ shows that short-distance factorization
holds, in analogy to the case of inclusive DIS.  QCD factorization
theorems~\cite{Collins:1996fb} state that in the limit of large $Q^2$
(at fixed Bjorken variable $x_B$ and fixed $t$) 
the Compton amplitude factorizes into a hard-scattering
subprocess and a hadronic matrix element describing the emission and
reabsorption of a parton by the proton target (see
Fig.~\ref{meson-fact}a).  As shown in Fig.~\ref{meson-fact}b, the
analogous result for exclusive meson production involves in addition
the quark-antiquark distribution amplitude of the meson (often termed
the meson wave function) and thus a further piece of non-perturbative
input.

The hadronic matrix elements appearing in the factorization formulae
for exclusive processes would be the usual PDFs if the proton had the same 
momentum in the initial and final state.
  Since this is not the case, they are more general
  functions taking into account the momentum difference between the
  initial and final state proton (or, equivalently, between the emitted
  and reabsorbed parton).  
These ``generalized parton distributions'' (GPDs) depend on two
independent longitudinal momentum fractions instead of a single one
(compare Figs.~\ref{compton-fact}a and \ref{meson-fact}a),
  on the transverse momentum transferred to the proton (whose
  square is $-t$ to a good approximation at high energy), and on the
  scale at which the partons are probed.  The scale dependence of the GPDs
  is governed by a generalization of the DGLAP equations.  
The dependence on the difference of the longitudinal momenta (often
called ``skewness'') contains information on correlations between
parton momenta in the proton wave function.  It can be neglected in
the approximation of leading $\log x$ (then the GPDs at $t=0$ reduce
to the usual PDFs as anticipated above), but it is numerically 
important in typical HERA kinematics.  The dependence on $t$ allows for a 
very intuitive interpretation if a Fourier transformation is performed 
with respect to the transverse momentum transfer. We then obtain
distributions depending on the impact parameter of the partons, which
describe the two-dimensional distribution of the struck parton in the
transverse plane, and on its longitudinal momentum fraction in
the proton.  The $t$ dependence of exclusive diffractive processes
thus provides unique information beyond the longitudinal momentum
spectra encoded in the usual parton densities.  The study of the 
generalized parton distributions is a prime reason to measure DVCS
and exclusive meson production in $ep$ scattering.  Detailed
discussions and references can be found in the recent
reviews~\cite{Diehl:2003ny,Belitsky:2005qn}.

\begin{figure}[p]
\begin{center}
\includegraphics[width=0.5\textwidth,bb=7 77 520 520]{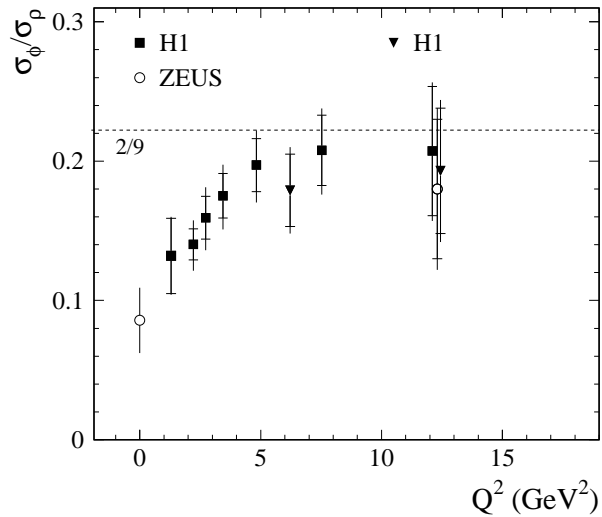}
\caption{\label{phi-to-rho} The ratio of cross sections for $\gamma^*
  p\to\phi p$ and $\gamma^* p\to\rho p$ as a function of the photon
  virtuality (from \protect\cite{Adloff:2000nx}).}
\end{center}
\end{figure}

An observable illustrating the short-distance factorization in meson
production at high $Q^2$ is the ratio of the $\phi$ and $\rho$
production cross sections, shown in Fig.~\ref{phi-to-rho}.  At large
$Q^2$ the process is described in terms of a light quark coupling to the 
photon
and of the generalized gluon distribution.  Using approximate flavor
SU(3) symmetry between the $\rho$ and $\phi$ wave functions, the only
difference between the two channels is then due to different quark
charge and isospin factors, which result in a cross section ratio of
$2/9$.

%%%%%%%%%%%%%%%%%%%%%%%%%%%%%%%%%%%%%%%%%%%%%%%%%%%%%%%%%%%%%

\subsection{High-energy factorization and the dipole picture}
\label{sec:dipole-pic}

So far we have discussed the description of hard exclusive diffraction
within short-distance, or collinear factorization.  A different type
of factorization is high-energy, or $k_t$ factorization, which is
based on the BFKL formalism.  Here the usual or generalized gluon
distribution appearing in the factorization formulae depends
explicitly on the transverse momentum $k_t$ of the emitted gluon.  In
collinear factorization, this $k_t$ is integrated over in the parton
distributions and set to zero when calculating the hard-scattering
process (the partons are thus approximated as ``collinear'' with
their parent hadron).  Likewise, the meson wave functions appearing in
$k_t$ factorization explicitly depend on the relative transverse
momentum between the quark and antiquark in the meson, whereas this is
integrated over in the quark-antiquark distribution amplitudes (cf. 
Sect.~\ref{exclusive}) of the collinear
factorization formalism.  Only gluon distributions appear in $k_t$
factorization, whereas collinear factorization formulae involve both
quark and gluon distributions (see e.g.\ Sects.~8.1 and 8.2 in
\cite{Diehl:2003ny} for a discussion of this difference).  We note
that other factorization schemes have been developed, which combine
features of the collinear and $k_t$ factorization formalisms.

The two different types of factorization implement different ways of
separating different parts of the dynamics in a scattering process.
The building blocks in a short-distance factorization formula correspond 
to
  either small or large particle virtuality (or equivalently to
  small or large transverse momentum), whereas the separation criterion
   in high-energy
factorization is the particle rapidity.  Collinear and $k_t$
factorization are based on taking different limits: in the former case
the limit of large $Q^2$ at fixed $x_B$ and in the latter case the
limit of small $x_B$ at fixed $Q^2$ (which must however be large
enough to justify the use of QCD perturbation theory).  In the common
limit of large $Q^2$ and small $x_B$ the two schemes give coinciding
results.  Instead of large $Q^2$ one can also take a large quark mass
in the limits just discussed.

\begin{figure}[b]
\begin{center}
\includegraphics[width=14cm]{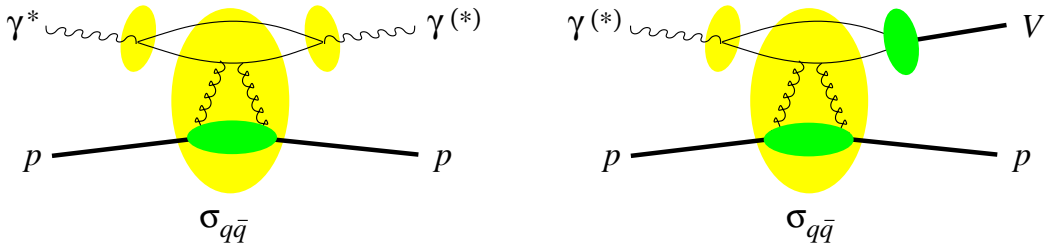}
\caption{\label{dipole-pic} The dipole representation of the
amplitudes for Compton scattering (a) and for meson production (b),
corresponding to the graphs in Figs.~\protect\ref{compton-fact}a and
\protect\ref{meson-fact}.}
\end{center}
\end{figure}

A far-reaching representation of high-energy dynamics can be obtained
by casting the results of $k_t$ factorization into a particular form.
The different building blocks in the graphs for Compton scattering and
meson production in Figs.~\ref{compton-fact}a and~\ref{meson-fact} can
be rearranged as shown in Fig.~\ref{dipole-pic}.  The result admits a
very intuitive interpretation in a reference frame where the photon
carries large momentum (this may be the proton rest frame but also a
frame where the proton moves fast, see Fig.~\ref{time-ordering}):
the initial photon splits into a
quark-antiquark pair, which scatters on the proton and finally forms a
photon or meson again.  This is the picture we have already appealed
to in Sect.~\ref{diff-at-hera}.

In addition, one can perform a Fourier transformation and trade the
relative transverse momentum between quark and antiquark for their
transverse distance $r$, which is conserved in the scattering on the
target.  The quark-antiquark pair acts as a color dipole, and its
scattering on the proton is described by a ``dipole cross section''
$\sigma_{q\bar{q}}$ depending on $r$ and on $\xpom$ (or on $x_B$ in the
case of inclusive DIS).  The wave functions of the photon and the meson 
depend on $r$ after Fourier transformation, and at small $r$ the photon 
wave function is perturbatively calculable.  Typical values of $r$ in a
scattering process are determined by the inverse of the hard momentum
scale, i.e.\ $r \sim (Q^2+M_V^2)^{-1/2}$.  An important result of
high-energy factorization is the relation
\begin{equation}
  \label{dip-gluon}
\sigma_{q\bar{q}}(r, x) \propto r^2 x g(x) 
\end{equation}
at small $r$, where we have replaced the generalized gluon
distribution by the usual 
one in the spirit of the leading $\log x$ approximation.  A more
precise version of the relation (\ref{dip-gluon}) involves the $k_t$
dependent gluon distribution.  The dipole cross section
vanishes at $r=0$ in accordance with the phenomenon of ``color
transparency'': a hadron becomes more and more transparent for a color
dipole of decreasing size.

The scope of the dipole picture is wider than we have presented so
far.  It is tempting to apply it outside the region where it can be
derived in perturbation theory, by modeling the dipole cross section
and the photon wave function at large distance $r$.  This has been very
been fruitful in phenomenology, as we will see in the next section.

\begin{figure}[tb]
\begin{center}
\includegraphics[width=9cm]{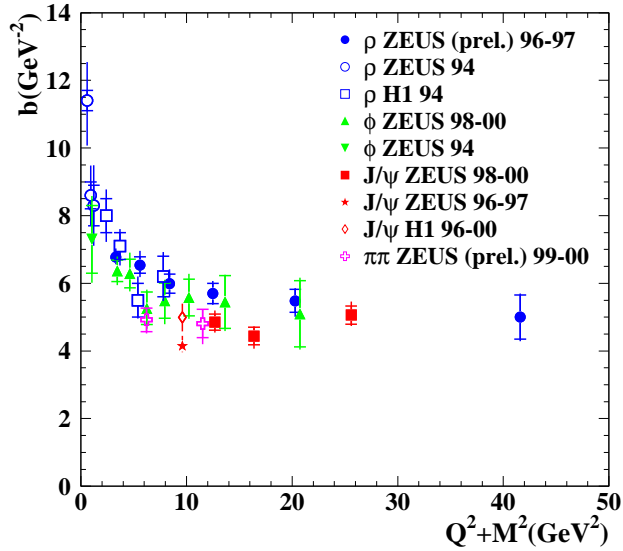}
\caption{\label{b-slopes} The logarithmic slope of the $t$ dependence
  at $t=0$ for different meson production channels, as well as for
  non-resonant dipion production.}
\end{center}
\end{figure}

The dipole picture is well suited to understand the $t$ dependence of 
exclusive processes, 
parameterized as $d\sigma/dt \propto
\exp(-b|t|)$ at small $t$.  Figure~\ref{b-slopes} shows that $b$
decreases with increasing scale $Q^2+M_V^2$ and at high scales becomes
independent of the produced meson.  A Fourier transform from momentum
to impact parameter space readily shows that $b$ is related to the
typical transverse distance between the colliding objects, as
anticipated by the analogy with optical diffraction in
Sect.~\ref{optics}.  At high scale, the $q\bar{q}$ dipole is almost
pointlike, and the $t$ dependence of the cross section is controlled by
the $t$ dependence of the generalized gluon distribution, or in
physical terms, by the transverse extension of the proton.  As
the scale decreases, the dipole acquires a size of its own, and in the
case of $\rho$ or $\phi$ photoproduction, the values of
$b$ reflect the fact that the two colliding objects are of typical
hadronic dimensions; similar values would be obtained in elastic
meson-proton scattering.

%%%%%%%%%%%%%%%%%%%%%%%%%%%%%%%%%%%%%%%%%%%%%%%%%%%%%%%%%%%%%

\subsection{Exclusive diffraction in hadron-hadron collisions}

\begin{figure}[htb]
\begin{center}
\includegraphics[width=7cm]{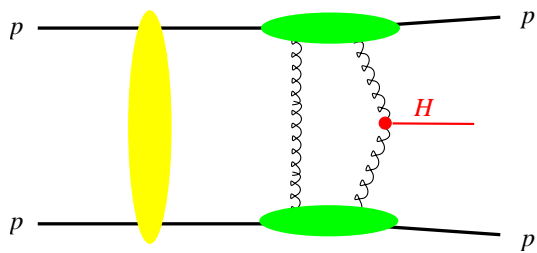}
\caption{\label{exclusive-higgs} Graph for the exclusive production of
  a Higgs boson in $pp$ scattering.  The horizontal blobs indicate
  generalized gluon distributions, and the vertical blob represents
  secondary interactions between the projectiles 
  (cf.\ Fig.~\protect\ref{fact-break}).}
\end{center}
\end{figure}

The concepts we have introduced to describe exclusive diffraction can
be taken over to $pp$ or $p\bar{p}$ scattering, although further
complications appear in these processes. A most notable reaction is
exclusive production of a Higgs boson, $pp\to pHp$, sketched in
Fig.~\ref{exclusive-higgs}.  The generalized gluon distribution is a
central input in this description.  The physics interest, theory
description, and prospects to measure this process at the LHC have
been discussed in detail at this workshop
\cite{procs-higgs-exp,procs-higgs-theo}.  A major
challenge in the description of this process is to account for
secondary interactions between spectator partons of the two
projectiles, which can produce extra particles in the final state and
hence destroy the rapidity gaps between the Higgs and final-state
protons -- the very same mechanism we discussed in
Sect.~\ref{survival}.

%------------------------------------------------------------------------------

\section{Parton saturation}

\begin{figure}[th]
\begin{center}
\includegraphics[width=7.2cm,angle=-90,clip=true]{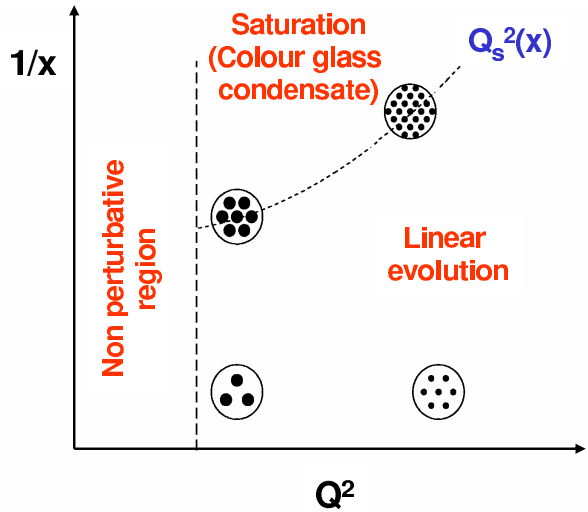}
\caption{\label{gluon-dens} Schematic view of the density of
  gluons in the transverse plane, as a function of the momentum fraction
  $x$ and the resolution scale $Q^2$.  Above the line given by $Q_s^2(x)$,
  saturation effects set in.}
\end{center}
\end{figure}

We have seen that diffraction involves scattering on
small-$x$ gluons in the proton.  Consider the density in the
transverse plane of gluons with longitudinal momentum fraction $x$
that are resolved in a process with hard scale $Q^2$.  One can think
of $1/Q$ as the ``transverse size'' of these gluons as seen by the
probe.  The number density of gluons at given $x$ increases with
increasing $Q^2$, as described by DGLAP evolution (see
Fig.~\ref{gluon-dens}).  According to the BFKL evolution equation it
also increases at given $Q^2$ when $x$ becomes smaller, so that the
gluons become more and more densely packed.  At some point, they will
start to overlap and thus reinteract and screen each other.  One then
enters a regime where the density of partons saturates and where the
linear DGLAP and BFKL evolution equations cease to be valid.  If $Q^2$
is large enough to have a small coupling $\alpha_s$, we have a theory
of this non-linear regime called ``color glass condensate'', see
e.g.~\cite{Weigert:2005us}.  To quantify the onset of non-linear
effects, one introduces a saturation scale $Q^2_s$ depending on $x$,
such that for $Q^2 < Q^2_s(x)$ these effects become important.  For
smaller values of $x$, the parton density in the target proton is
higher, and saturation sets in at larger values of $Q^2$ as
illustrated in Fig.~\ref{gluon-dens}.

The dipole picture we introduced in Sect.~\ref{sec:dipole-pic} is well
suited for the theoretical description of saturation effects.  When
such effects are important, the relation (\ref{dip-gluon}) between
dipole cross section and gluon distribution ceases to be valid; in
fact the gluon distribution itself is then no longer an adequate
quantity to describe the dynamics of a scattering process.  In a
certain approximation, the evolution of the dipole cross section with
$x$ is described by the Balitsky-Kovchegov
equation~\cite{Balitsky:1995ub}, which supplements the BFKL equation
with a non-linear term taming the growth of the dipole cross section
with decreasing $x$.

\begin{figure}[p]
\begin{center}
\includegraphics[width=0.55\textwidth]{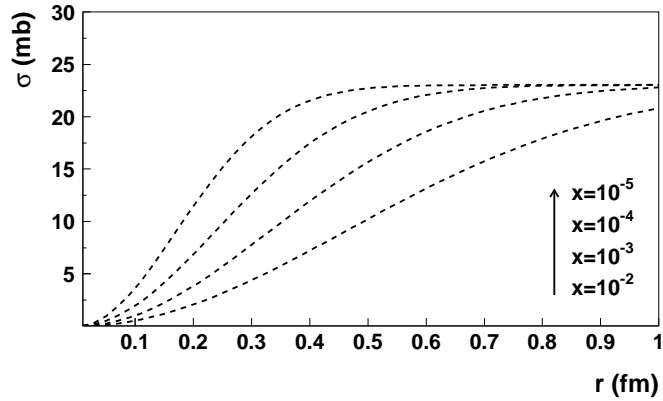}
\caption{\label{gbw-xsect} The dipole cross section $\sigma_{q\bar{q}}$ in 
the Golec-Biernat W\"usthoff model as a function of dipole size $r$ for
  different $x$ (from~\protect\cite{Golec-Biernat:2002bj}).}
\end{center}
\end{figure}

Essential features of the saturation phenomenon are captured in a
phenomenological model for the dipole cross section, originally
proposed by Golec-Biernat and W\"usthoff, see
\cite{Golec-Biernat:1999qd,Golec-Biernat:2002bj}.
Figure~\ref{gbw-xsect} shows $\sigma_{q\bar{q}}$ as a function of $r$ at
given $x$ in this model. The dipole size $r$ now plays the role of
$1/Q$ in our discussion above.  At small $r$ the cross section rises
following the relation $\sigma_{q\bar{q}}(r, x) \propto r^2 x g(x)$. 
At some value $R_s(x)$ of $r$, the dipole cross section is so large
  that this relation ceases to be valid, and $\sigma_{q\bar{q}}$ starts to
  deviate from the quadratic behavior in $r$.
  As $r$ continues to increase, $\sigma_{q\bar{q}}$ eventually saturates 
  at a value typical of a meson-proton cross section.
In terms of the
saturation scale introduced above, $R_s(x) = 1/Q_s(x)$.  For
smaller values of $x$, the initial growth of $\sigma_{q\bar{q}}$ with $r$ 
is stronger because the gluon distribution is larger.  The target is thus
more opaque and as a consequence saturation sets in at lower $r$.

\begin{figure}[p]
\begin{center}
\includegraphics[width=0.74\textwidth,bb=0 15 460 560]{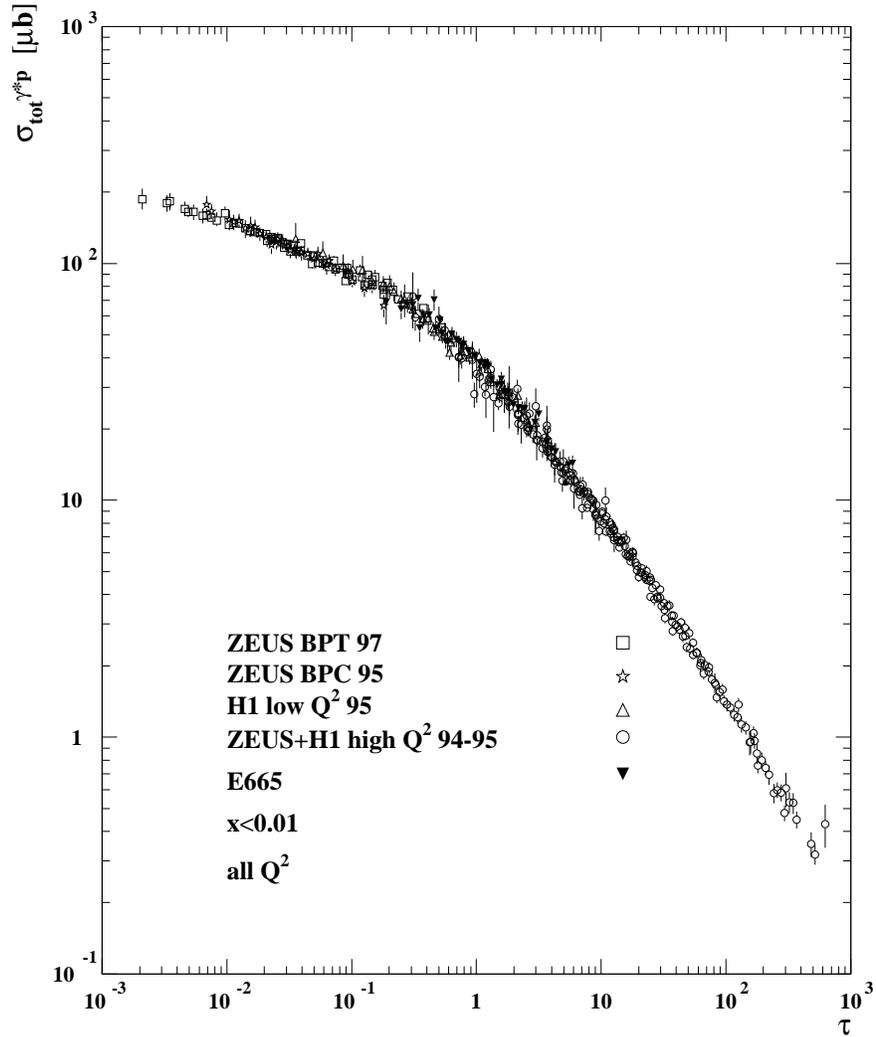}
\caption{\label{geo-scaling} Geometric scaling of the $\gamma^* p$
  cross section in a single variable $\tau =Q^2 /Q^2_s(x_B)$, as
  determined in \protect\cite{Stasto:2000er}.  The $Q^2$ of these data
  ranges from $0.045$ to $450$~GeV$^2$.}
\end{center}
\end{figure}

A striking feature found both in this phenomenological
model~\cite{Stasto:2000er} and in the solutions of the
Balitsky-Kovchegov equation (see e.g.~\cite{Golec-Biernat:2001if}) is
that the total $\gamma^* p$ cross section only depends on $Q^2$ and
$x_B$ through a single variable $\tau= Q^2 /Q^2_s(x_B)$.  This
property, referred to as geometric scaling, is well satisfied by the
data at small $x_B$ (see Fig.~\ref{geo-scaling}) and is an important
piece of evidence that saturation effects are visible in these data.
Phenomenological estimates find $Q^2_s$ of the order 1~GeV$^2$ for
$x_B$ around $10^{-3}$ to $10^{-4}$.

\begin{figure}[ht]
\begin{center}
\includegraphics[width=0.7\textwidth]{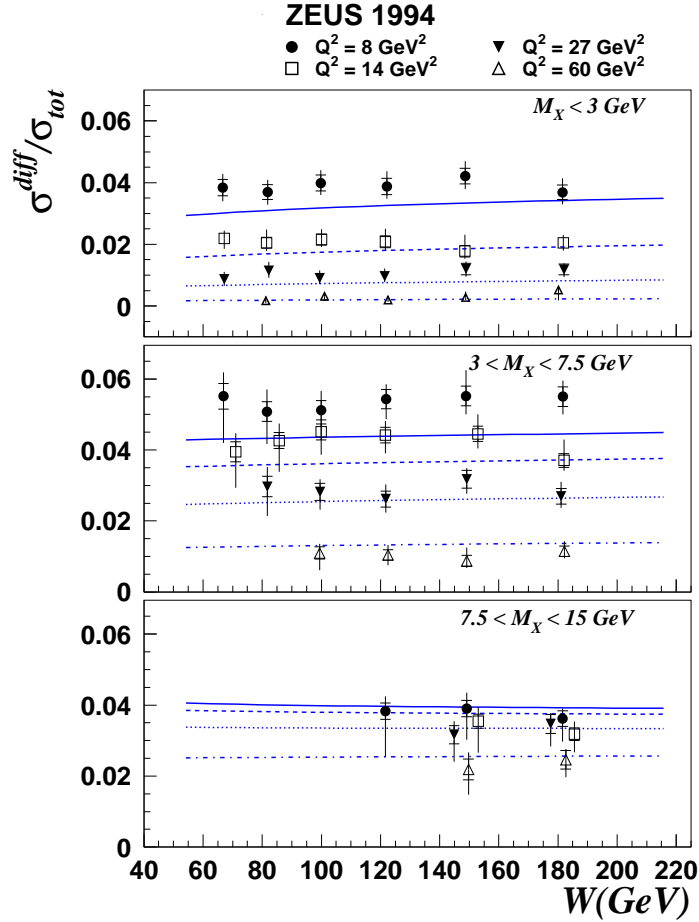}
\caption{\label{gbw} Data on the ratio of diffractive and total
  $\gamma^*p$ cross sections compared with the result of the
  Golec-Biernat W\"usthoff model (from
  \protect\cite{Golec-Biernat:1999qd}).}
\end{center}
\end{figure}

The dipole formulation is suitable to describe not only exclusive
  processes and inclusive DIS, but also inclusive diffraction
  $\gamma^* p\to X p$.
For a diffractive final state $X=q\bar{q}$ at
parton level, the theory description is very similar to the one for
deeply virtual Compton scattering, with the wave function for the
final state photon replaced by plane waves for the produced $q\bar{q}$
pair. The inclusion of the case $X=q\bar{q}g$ requires further
approximations~\cite{Golec-Biernat:1999qd} but is phenomenologically
indispensable for moderate to small $\beta$.  
Experimentally, one observes a very similar energy dependence of the
 inclusive diffractive and the total cross section in $\gamma^* p$
 collisions at given $Q^2$ (see Fig.~\ref{gbw}).  The saturation
 mechanism implemented in the Golec-Biernat W\"usthoff model provides
 a simple explanation of this finding. To explain this aspect of
the data is non-trivial.  For instance, in the description based on
collinear factorization, the energy dependence of the inclusive and
diffractive cross sections is controlled by the $x$ dependence of the
ordinary and the diffractive parton densities.  This $x$ dependence is
not predicted by the theory.

The description of saturation effects in $pp$, $pA$ and $AA$
collisions requires the full theory of the color glass condensate,
which contains concepts going beyond the dipole formulation discussed
here and is e.g.\ presented in \cite{Weigert:2005us}.  We remark
however that estimates of the saturation scale $Q_s^2(x)$ from HERA
data can be used to describe features of the recent data from RHIC
\cite{Kharzeev:2001gp}.

%------------------------------------------------------------------------------

\section{A short summary}

Many aspects of diffraction in $ep$ collisions can be successfully
described in QCD if a hard scale is present.  A key to this success
are factorization theorems, which render parts of the dynamics
accessible to calculation in perturbation theory.  The remaining
non-perturbative quantities, namely diffractive PDFs and generalized
parton distributions, can be extracted from measurements and
contain specific information about small-$x$ partons in the proton that 
can only be obtained in diffractive processes.
To describe hard diffractive hadron-hadron collisions is more
challenging since factorization is broken by rescattering between
spectator partons.  These rescattering effects are of interest in
their own right because of their intimate relation with multiple
scattering effects, which at LHC energies are expected to be crucial
for understanding the structure of events in hard collisions.  A
combination of data on inclusive and diffractive $ep$ scattering hints
at the onset of parton saturation at HERA, and the phenomenology
developed there is a helpful step towards understanding high-density
effects in hadron-hadron collisions.

%------------------------------------------------------------------------------

\section*{Acknowledgments}

It is a pleasure to thank our co-convenors and all participants for the
fruitful atmosphere in the working group on diffraction, and A.~De Roeck
and H.~Jung for their efforts in organizing this workshop. We are indebted
to A.~Proskuryakov for Figs.~\ref{xlt_spectrum} and~\ref{fig-cdf}, to
P.~Fleischmann for Fig.~\ref{vmXsec}, and to A.~Bonato, K.~Borras, 
A.~Bruni, J.~Forshaw, M.~Grothe, H.~Jung, L.~Motyka, M.~Ruspa and M.~Wing 
for valuable comments on the manuscript.

%------------------------------------------------------------------------------
%       Bibliography
%------------------------------------------------------------------------------
%\bibliographystyle{heralhc} 
%{\raggedright
%\bibliography{heralhc}
%}

\end{document}